\documentclass[a4paper,12pt]{article}
\usepackage{amssymb,amsmath,eepic,psfrag,epsfig}

\def\d{\mathrm{d}}
\def\H{\mathcal{H}}
\def\D{\mathcal{D}}
\def\V{\mathcal{V}}
\begin{document}
\title{The problem of uniqueness in the reduced description of adsorption on 
the wedge-shaped substrate}
\author{A.Bednorz and M.Napi\'orkowski,\\
Instytut Fizyki Teoretycznej, Uniwersytet Warszawski\\
00-681 Warszawa, Ho\.za 69, Poland}
\maketitle
\begin{abstract}
In the reduced one-dimensional description of the adsorption on the 
wedge-shaped substrate the mid-point interface height serves as the  
order parameter. We point at the ambiguity which appears in the 
transfer-matrix approach to this problem. We also propose how to avoid 
this problem by introducing the appropriate order parameter. \\ 

\end{abstract}
\newpage
\centerline{\bf {I. Introduction}}
\renewcommand{\theequation}{1.\arabic{equation}} 
\setcounter{equation}{0}
\vspace*{0.5cm}

One of the possible scenarios of adsorption on 
the wedge-shaped substrate, see Fig.1, proceeds is via the so-called 
critical filling transition [1-6]. In this transition the central part 
of the interface (separating the phases $\beta$ and 
$\alpha$) positioned above the edge of the wedge is shifted continuously 
to infinity while the parts of the interface 
corresponding to $|x| \rightarrow \infty$ remain pinned to the substrate. 
The filling transition takes place at the temperature $T_{\varphi}$ which 
depends on the wedge opening angle $2\varphi$ and which is smaller than 
the wetting temperature $T_{w}$ on the planar substrate. \\ 
The critical filling transition was analyzed recently \cite{Parry} via the 
transfer-matrix approach. Due to the strong anisotropy of the interfacial 
fluctuations the order parameter corresponding to the height of the 
interface $\ell(x,y)$ above the substrate $z=|x| \cot\varphi$ can be 
effectively replaced by the mid-point height $\ell(y)=\ell(0,y)$. The 
corresponding one-dimensional Hamiltonian has the following form
\cite{Parry,BN1}
\begin{equation}
H[\ell(y)]=\int\d y\H=\int\d y\frac{\sigma}{\alpha}\left[
\ell(y)\left(\frac{\d\ell}{\d y}\right)^2+(\Theta^2-\alpha^2)\ell(y)\right]\,,
\end{equation}
where $\sigma$ is the $\alpha$-$\beta$ surface tension and the planar 
substrate \emph{contact angle} $\Theta$ is defined via the Young equation. 
We consider very opened wedge and thus we have put $\alpha = \cos\varphi 
\approx \cot\varphi$. Actually the factor $\Theta^2-\alpha^2$ in Eq.(1.1) 
measures the dimensionless deviation from the filling temperature because 
$\Theta(T_{\varphi})=\alpha$. 
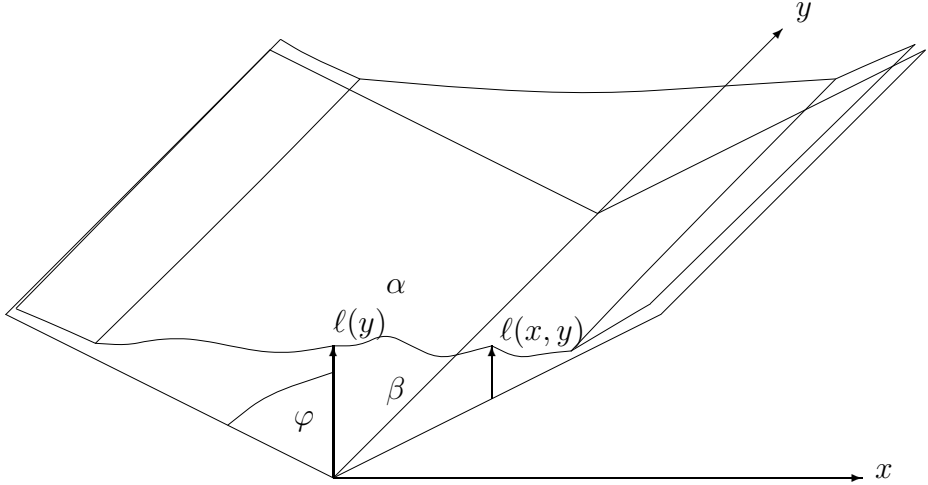
\begin{figure}
\begin{center}
\begin{picture}(300,200)
\put(100,0){\vector(1,0){200}}
\put(110,150){\spline(0,1)(45,-3)(90,-5)(135,-2)(180,1)}
\put(10,50){\spline(0,1)(50,50)(100,101)}
\put(10,50){\spline(0,1)(-20,10)(-30,14)}
\put(110,150){\spline(0,1)(-20,10)(-30,16)}
\put(-20,65){\spline(0,-1)(35,35)(100,101)}
\put(190,50){\spline(0,-2)(35,35)(100,101)}
\put(190,50){\spline(0,-2)(20,10)(30,16)}
\put(290,150){\spline(0,1)(20,10)(30,14)}
\put(220,65){\spline(0,1)(50,50)(100,99)}
\put(200,100){\vector(1,1){70}}

\put(100,50){\spline(0,0)(-10,-2)(-20,-3)(-30,-2)(-40,0)(-50,2)(-60,3)(-70,2)
(-80,0)(-90,1)}
\put(100,0){\vector(0,1){50}}

\put(160,30){\vector(0,1){20}}
\put(100,50){\spline(0,0)(10,0)(20,5)(30,0)(40,-5)(50,-3)(60,0)}
\put(160,50){\spline(0,0)(10,-5)(20,-3)(30,-2)}
\put(275,175){$y$}
\put(305,0){$x$}
\put(163,52){$\ell(x,y)$}
\put(100,55){$\ell(y)$}
\put(100,0){\spline(-40,20)(-25,32)(0,40)}
\put(85,20){$\varphi$}
\put(120,30){$\beta$}
\put(120,70){$\alpha$}

\path(-24,62)(76,162)
\path(100,0)(200,100)
\put(100,0){\line(-2,1){124}}
\put(100,0){\line(2,1){124}}
\put(224,62){\line(1,1){100}}
\put(200,100){\line(-2,1){124}}
\put(200,100){\line(2,1){124}}
\end{picture}
\end{center}
\caption{The wedge geometry and the fluctuating $\alpha-\beta$ 
interface}\label{kli}
\end{figure}
The above one-dimensional Hamiltonian can be further simplified by 
introducing the rescaled variables $Y$ and $L$ 
\begin{equation}
\alpha y=\Sigma^{-1/2}((\Theta/\alpha)^2-1)^{-3/4}Y,\,\,\ell=\Sigma^{-1/2}
((\Theta/\alpha)^2-1)^{-1/4}L\,,
\end{equation}
Then the Hamiltonian becomes free from any parameters and has the form 
\begin{equation}
H[L(Y)]=\int\d Y L\left[(L'(Y))^2+1\right]\,. 
\end{equation}
This scaling property leads straightforwardly to the critical behavior of
the mean mid-height $\langle\ell\rangle\sim(\Theta/\alpha-1)^{-1/4}$ and 
the correlation length $\xi_y\sim(\Theta/\alpha-1)^{-3/4}$ \cite{Parry}. 


\centerline{\bf {II. The propagator }}
\renewcommand{\theequation}{2.\arabic{equation}} 
\setcounter{equation}{0}
\vspace{0.5cm}

To solve the model described by the Hamiltonian in Eq.(1.3) one introduces 
the propagator \cite{Burk}
\begin{equation}
V(L_2,L_1,Y_2,Y_1)=\int\D L\exp(-H[L])|_{L(Y_1)=L_1}^{L(Y_2)=L_2}
\end{equation}
where the measure $\D L$ is given by $\D L=\prod_Y L^{1/2}(Y)\d L(Y)$. 
Actually it is the form of this measure which prohibits one from deriving 
the equation for the propagator in an unambiguous way. The problem 
encountered here is similar to the well known It\^o-Stratonovich dilemma 
in the theory of stochastic processes \cite{Kam}. \\ 

For $Y_2\,-\,Y_1=\Delta Y\ll 1$ the discretization schemes applied to 
Eq.(2.1) can be 
parametrized by two parameters $a$ and $b$ ($a,b\in[0,1]$). These two 
parameters reflect the freedom (or rather ambiguity) in: i) defining the 
measure because of the factor $L^{1/2}(Y)$ present in the measure 
$\prod_Y L^{1/2}(Y)\d L(Y)$, $a$; and ii) defining the discrete analogue of 
the term $L(Y)\,(dL(Y)/dY)^2$ present in the Hamiltonian, $b$. In each of 
these cases the factor $L^{1/2}$ can be split into two factors $L^{c/2}$ 
and $L^{(1-c)/2}$, $c=a,b$  attached to the left and to the right end of 
the segment $\Delta Y$, respectively. In this way one obtains 
\begin{equation}
V(L_2,L_1,\Delta Y)=L_2^{(1-a)/2}L_1^{a/2}\exp\left\{-[(1-b)L_2+bL_1]
\frac{(L_2-L_1)^2}{\Delta Y}-L_2\Delta Y\right\}.
\end{equation}
which leads to the following equation for the propagator (the Fokker-Planck 
equation) in the limit $\Delta Y \rightarrow 0$ 
\begin{equation}
\frac{\partial V}{\partial Y}=-L_2V+\frac{\partial^2 V}
{4L_2\partial L_2^2}-\frac{(3b-a)\partial V}{4L_2^2\partial L_2}
+\frac{(15b^2-6ab-a^2)V}{16L_2^3}.\label{eqel}
\end{equation}
We see that the form of this equation depends on the choice of parameters 
$a$ and $b$. If one insists that the propagator is symmetric, i.e. 
invariant upon interchanging $L_1$ and $L_2$ then one obtains the condition 
$3b-a=1$ which still leaves the equation for the propagator depending on 
one parameter. \\
The above ambiguity can be avoided by changing the variable in the 
one-dimensional Hamiltonian in Eq.(1). Instead of the variable $L$ one 
introduces the new order parameter $\eta\equiv 2L^{3/2}/3$ and the 
Hamiltonian takes the form 
\begin{equation}
H[\eta(Y)]=\int\d Y\left[(\eta'(y))^2+(3\eta/2)^{2/3}\right].
\end{equation}
The corresponding  propagator is defined as 
\begin{equation}
\V(\eta_2,\eta_1,Y_2,Y_1)=\int\D\eta\exp(-H[\eta])|_{\eta(Y_1)=
\eta_1}^{\eta(Y_2)=\eta_2}\,,
\end{equation}
where $\D\eta=\prod_Y\d\eta(Y)$.
Now the equation for the propagator is obtained unambiguously 
\begin{equation}
\frac{\partial\V}{\partial Y}=\frac{\partial^2\V}
{4\partial\eta_2^2}-(3\eta_2/2)^{2/3}\V\label{eqeta}\,.
\end{equation}
The propagators $\V(\eta_1,\eta_2,Y)$ and $V(L_1,L_2,Y)$ are related 
\begin{displaymath}
V(L_1,L_2,Y_1,Y_2)=(2/3)^{1/3}(\eta_1\eta_2)^{1/6}\V(\eta_1,\eta_2,Y_1,Y_2).
\end{displaymath} 
It is interesting to note that Eq.(\ref{eqel}) for the "symmetrical  
choice" $a=b=1/2$ is not equivalent to Eq.(\ref{eqeta}).\\

\centerline{\bf {III. The boundary condition}}
\renewcommand{\theequation}{3.\arabic{equation}} 
\setcounter{equation}{0}
\vspace{0.5cm}

The equation for the propagator must be supplemented by appropriate boundary 
conditions at $\eta=0$, i.e. at $L=0$. In this letter we follow \cite{Burk} 
and impose the following condition 
\begin{equation}
\left.\frac{\partial\V(\eta_2,
\eta_1,Y_2,Y_1)}{\partial\eta_2}\right|_{\eta_2=0} =a\V(0,\eta_1,Y_2,Y_1).
\end{equation}
This condition should not depend on $\Theta-\alpha$. Thus for the 
non-rescaled variable $\bar{\eta}$ defined as 
$\bar{\eta}=(\Theta/\alpha-1)^{-3/8}\eta$ one  must have
\begin{equation}
\partial_{\bar{\eta}_2}\ln\V|_{\bar{\eta}_2}=\bar{a}=\mathrm{const}.
\end{equation}
Therefore the parameter $a=\bar{a}(\Theta/\alpha-1)^{-3/8}$ tends to 
$\infty$ upon approaching the filling temperature from which one concludes 
that the correct boundary condition has the Dirichlet form 
\begin{equation}
\V(0,\eta_1,Y_2,Y_1)=0\,.
\end{equation}

\begin{figure}
\begin{center}
\psfrag{psi0}{${\scriptstyle\psi_0(\eta)}$}
\psfrag{psi1}{${\scriptstyle\psi_1(\eta)}$}
\psfrag{psi2}{${\scriptstyle\psi_2(\eta)}$}
\psfrag{psi3}{${\scriptstyle\psi_3(\eta)}$}
\psfrag{eta}{${\scriptstyle\eta}$}
\epsfxsize=5cm
\epsffile{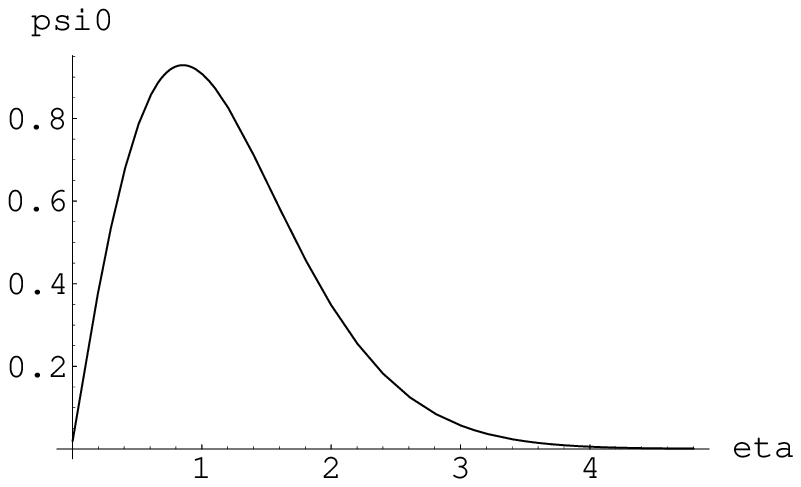}
\epsfxsize=5cm
\epsffile{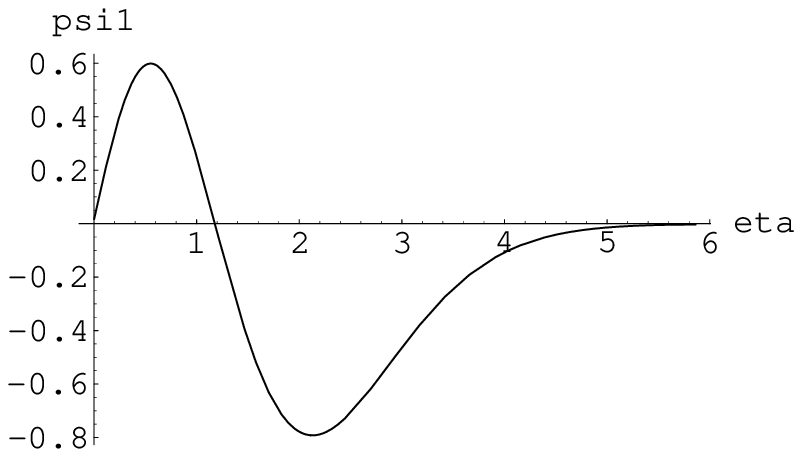}\\
\epsfxsize=5cm
\epsffile{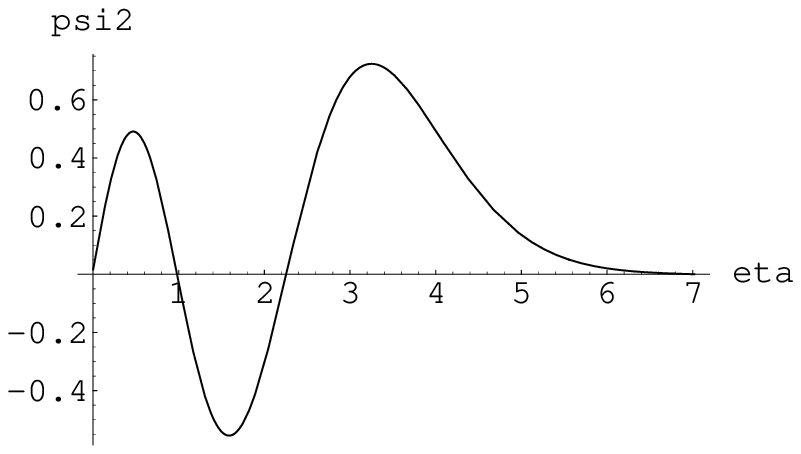}
\epsfxsize=5cm
\epsffile{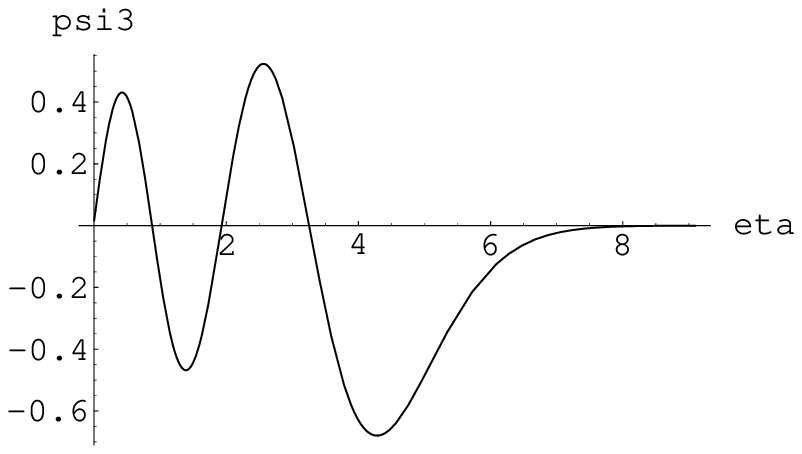}
\end{center}
\caption{The eigenfunctions $\psi_0$, $\psi_1$, $\psi_2$ and $\psi_3$.}  
\label{psss}
\end{figure}

In order to find the propagator explicitly we express it
by eigenvalues $E_n$ and eigenfunctions $\psi_n$ of the equation
\begin{equation}
\left[-E_n+(3\eta/2)^\frac{2}{3}-\partial^2_\eta/4
\right]\psi_n(\eta)=0
\end{equation}
with  boundary condition $\psi_n(0)=0$
Then the propagator is written is the from
\begin{equation}
\V(\eta_2,\eta_1,Y_2,Y_1)=\sum_n\psi_n(\eta_2)\psi_n(\eta_1)
e^{-E_n (Y_2-Y_1)}\,.
\end{equation}
The first four eigenvalues are $E_0\approx 1.75137$, $E_1\approx 2.65289$,
$E_2\approx 3.32079$ $E_3\approx 3.87586$ and the corresponding
eigenfunctions  obtained numerically \cite{hep} are shown on the
Fig. \ref{psss}.

To calculate physical quantities one needs the multipoint probability 
distribution $p(Y_0,\eta_0,\dots,Y_k,\eta_k)$. This distribution can be 
expressed as the product of propagators 
\begin{equation}
p(Y_0,\eta_0,\dots,Y_k,\eta_k)=\frac{\prod_{i=-1}^k\V(\eta_{i+1},\eta_i,
Y_{i+1},Y_i)}{\V(\eta_{k+1},\eta_{-1},Y_{k+1},Y_{-1})}\,.
\end{equation}
where $(Y_{-1},\eta_{-1})$ and $(Y_{k+1},\eta_{k+1})$ are coordinates of the
boondary conditions.\\

\centerline{\bf {IV. The conclusions}}
\renewcommand{\theequation}{4.\arabic{equation}} 
\setcounter{equation}{0}
\vspace{0.5cm}

We have pointed out that although the transfer-matrix method seems to 
be applicable rather straightforwardly to the effective one-dimensional 
Hamiltonian describing the critical fluctuations at the filling transition 
one is still left with the problem of non-unique way of discretizing this 
problem. Thus the analogue of the It\^o-Stratonovich dilemma appears in 
the transfer-matrix analysis of the critical interfacial fluctuations in the 
presence of non-planar substrate. In order to avoid this problem we propose 
to find first the right order parameter and the corresponding space of 
functional integration. We show how such a choice leads to the disappearance 
of the ambiguity upon the discretization of the problem. \\

The above considerations show that in order to get the hint about the right 
form of the  Fokker-Planck equation one should go back to the complete 
two-dimensional description and from there deduce the correct values of 
$a$ and $b$. 
Eq.(2.3) becomes equivalent to Eq.(2.6) if the term  $\epsilon\V/4$ is added 
to the rhs of Eq.(2.3), where the coefficient $\epsilon$ depends on the 
parameters  $a$ and $b$. For "symmetrical choice" $\epsilon=-1/36$. We 
suspect that $\epsilon$ is in fact nonzero and finding its right value 
remains the challenge. \\

\end{document}